\documentclass{ws-ijmpd}
\usepackage[super,compress]{cite}
\usepackage[breaklinks]{hyperref}
\hypersetup{colorlinks,urlcolor=black,citecolor=black,linkcolor=black,filecolor=black}
\usepackage{breakurl}

\usepackage{amsmath}
\usepackage{amssymb}
\usepackage{aas-macros}
\usepackage{url}

\begin{document}

\markboth{M.Yu. Piotrovich, S.D. Buliga \& T.M. Natsvlishvili}
{Estimating the spins of supermassive black holes in distant ultraluminous quasars}

%
\catchline{}{}{}{}{}
%

\title{ESTIMATING THE SPINS OF SUPERMASSIVE BLACK HOLES IN DISTANT ULTRALUMINOUS QUASARS\\
\medskip {\small \it Accepted for publication in International Journal of Modern Physics D}}

\author{MIKHAIL YU. PIOTROVICH}

\address{Central astronomical observatory at Pulkovo\\ 196140 St.-Petersburg, Russia\\ mpiotrovich@mail.ru}

\author{STANISLAVA D. BULIGA}

\address{Central astronomical observatory at Pulkovo\\ 196140 St.-Petersburg, Russia}

\author{TINATIN M. NATSVLISHVILI}

\address{Central astronomical observatory at Pulkovo\\ 196140 St.-Petersburg, Russia}

\maketitle

\begin{history}
\received{Day Month Year}
\revised{Day Month Year}
\end{history}

\begin{abstract}
We estimated spin, inclination angle and corresponding SMBH mass values for sample of extremely distant ($6 < z < 7.5$) ultraluminous quasars. The estimated spin values are on average greater that 0.9 and the spin distribution has a characteristic appearance, similar to ones obtained for other types of AGNs and quasars. The dependence of estimated parameters on each other shows strong correlations between them, from which we can assume that in this early quasars the growth of SMBHs mass should occur mainly due to disk accretion with high accretion rate, which very effectively increases the spin.
\end{abstract}


\ccode{PACS numbers: 97.10.Gz, 98.54.Aj}

\section{Introduction}

Quasars are the most luminous non-transient sources in the Universe and are observed down to redshifts $z \approx 7.6$  \cite{wang21}. They are a valuable source of information, providing key studies of galaxy evolution and cosmology throughout cosmic history on three important spatial scales. First, quasars are fundamental for studying the physics of AGN activity on scales $<$ 1 kpc, accretion and growth of black holes. Second, quasars are key to understanding the structure, growth and extinction of massive galaxies on the scale of 1-10 kpc. Finally, quasars can be used to study the growth of early large-scale structures ($<$ Mpc), and provide opportunities to study the properties of the intergalactic medium, including the epoch of reionization, chemical enrichment of the intergalactic medium, the distribution of baryons and the evolution of the ionization state of intergalactic medium.

To date, about 300 spectroscopically confirmed quasars have been detected at $6 < z < 7.5$ during the epoch of reionization \cite{fan23,mazzucchelli17}. These quasars are generated by accretion of SMBHs with masses of $10^{8}-10^{10} M_\odot$ and bolometric luminosity $L_\text{bol} = 10^{46}-10^{48}\,{\rm erg/s}$ \cite{mazzucchelli17,shen19,mazzucchelli23,dodorico23}. The very existence of SMBH masses of the order of $10^9 M_{\odot}$ at the reionization epoch creates serious problems for theoretical models to explain how these systems formed in less than 1 billion years \cite{volonteri10,johnson16}. Various scenarios are presented in the literature explaining the existence of SMBHs with masses of $\approx 10^{9} M_{\odot}$ at $z > 6$. Among the modern theoretical scenarios, one can note: a) direct collapse models, which allow accelerating the growth of SMBHs from massive BH seeds $M_\text{BH}^\text{seed} > 10^{3-4} M_{\odot}$, which can then reach a mass of a billion solar masses upon accretion to the Eddington limit \cite{begelman06,schauer17,dayal19}; b) a scenario of short and discontinuous phases of super-Eddington accretion, allowing the mass to increase by three orders of magnitude over a few million years from a BH seed of lower mass $M_\text{BH}^\text{seed} \approx 100 M_{\odot}$ \cite{tanaka09,inayoshi20}; c) the presence of radiation-inefficient supercritical accretion of stellar-mass black holes embedded in gaseous circumnuclear disks, which are expected to exist in the nuclei of high-redshift galaxies \cite{lupi16,trakhtenbrot17,davies19}.

In connection with all this, it is of interest to try to determine such an important physical parameter of the SMBHs as spin. So, in this paper, we estimate the spins and inclination angles of a sample of distant ultraluminous quasars at $6 < z < 7.5$.

\section{Analysis of initial data}

\begin{table}[ht!]
\tbl{Parameters of objects of study\cite{zappacosta23}: cosmological redshift $z$, bolometric luminosity $L_\text{bol}$, Eddington ratio $l_\text{E}$ and SMBH mass $M_\text{BH}$.}
{\begin{tabular}{lcccc}
\toprule
Object & $z$ & $\log{L_\text{bol}}$ & $\log{l_\text{E}}$ & $\log{\frac{M_\text{BH}}{M_\odot}}$\\
\colrule
 ULAS  J1342+0928 &  7.541$\pm$0.005 & 47.19$\pm$0.10 &  0.19$\pm$0.10 &  8.90$\pm$0.10 \\
  QSO  J1007+2115 &  7.494$\pm$0.005 & 47.30$\pm$0.10 &  0.02$\pm$0.10 &  9.18$\pm$0.10 \\
 ULAS  J1120+0641 &  7.087$\pm$0.005 & 47.30$\pm$0.10 & -0.21$\pm$0.10 &  9.41$\pm$0.10 \\
 DELS  J0038-1527 &  7.021$\pm$0.005 & 47.36$\pm$0.10 &  0.12$\pm$0.10 &  9.14$\pm$0.10 \\
  DES  J0252-0503 &  6.990$\pm$0.005 & 47.12$\pm$0.10 & -0.13$\pm$0.10 &  9.15$\pm$0.10 \\
 VDES  J0020-3653 &  6.834$\pm$0.005 & 47.16$\pm$0.10 & -0.18$\pm$0.10 &  9.24$\pm$0.10 \\
  VHS  J0411-0907 &  6.824$\pm$0.005 & 47.31$\pm$0.10 &  0.41$\pm$0.10 &  8.80$\pm$0.10 \\
 VDES  J0244-5008 &  6.724$\pm$0.005 & 47.19$\pm$0.10 &  0.01$\pm$0.10 &  9.08$\pm$0.10 \\
  PSO J231.6-20.8 &  6.587$\pm$0.005 & 47.31$\pm$0.10 & -0.29$\pm$0.10 &  9.50$\pm$0.10 \\
  PSO J036.5+03.0 &  6.533$\pm$0.005 & 47.33$\pm$0.10 & -0.26$\pm$0.10 &  9.49$\pm$0.10 \\
 VDES  J0224-4711 &  6.526$\pm$0.005 & 47.53$\pm$0.10 &  0.07$\pm$0.10 &  9.36$\pm$0.10 \\
  PSO     J011+09 &  6.444$\pm$0.005 & 47.12$\pm$0.10 & -0.13$\pm$0.10 &  9.15$\pm$0.10 \\
 SDSS  J1148+5251 &  6.422$\pm$0.005 & 47.57$\pm$0.10 & -0.27$\pm$0.10 &  9.74$\pm$0.10 \\
  PSO J083.8+11.8 &  6.346$\pm$0.005 & 47.16$\pm$0.10 & -0.26$\pm$0.10 &  9.32$\pm$0.10 \\
 SDSS  J0100+2802 &  6.300$\pm$0.005 & 48.24$\pm$0.10 &  0.10$\pm$0.10 & 10.04$\pm$0.10 \\
ATLAS     J025-33 &  6.294$\pm$0.005 & 47.39$\pm$0.10 & -0.14$\pm$0.10 &  9.57$\pm$0.10 \\
CFHQS  J0050+3445 &  6.246$\pm$0.005 & 47.29$\pm$0.10 & -0.49$\pm$0.10 &  9.68$\pm$0.10 \\
ATLAS     J029-36 &  6.027$\pm$0.005 & 47.39$\pm$0.10 & -0.52$\pm$0.10 &  9.82$\pm$0.10 \\
\botrule
\end{tabular} \label{table_01}}
\end{table}

\begin{figure}[ht!]
\centering
\includegraphics[bb= 60 25 715 530, clip, width=0.7\linewidth]{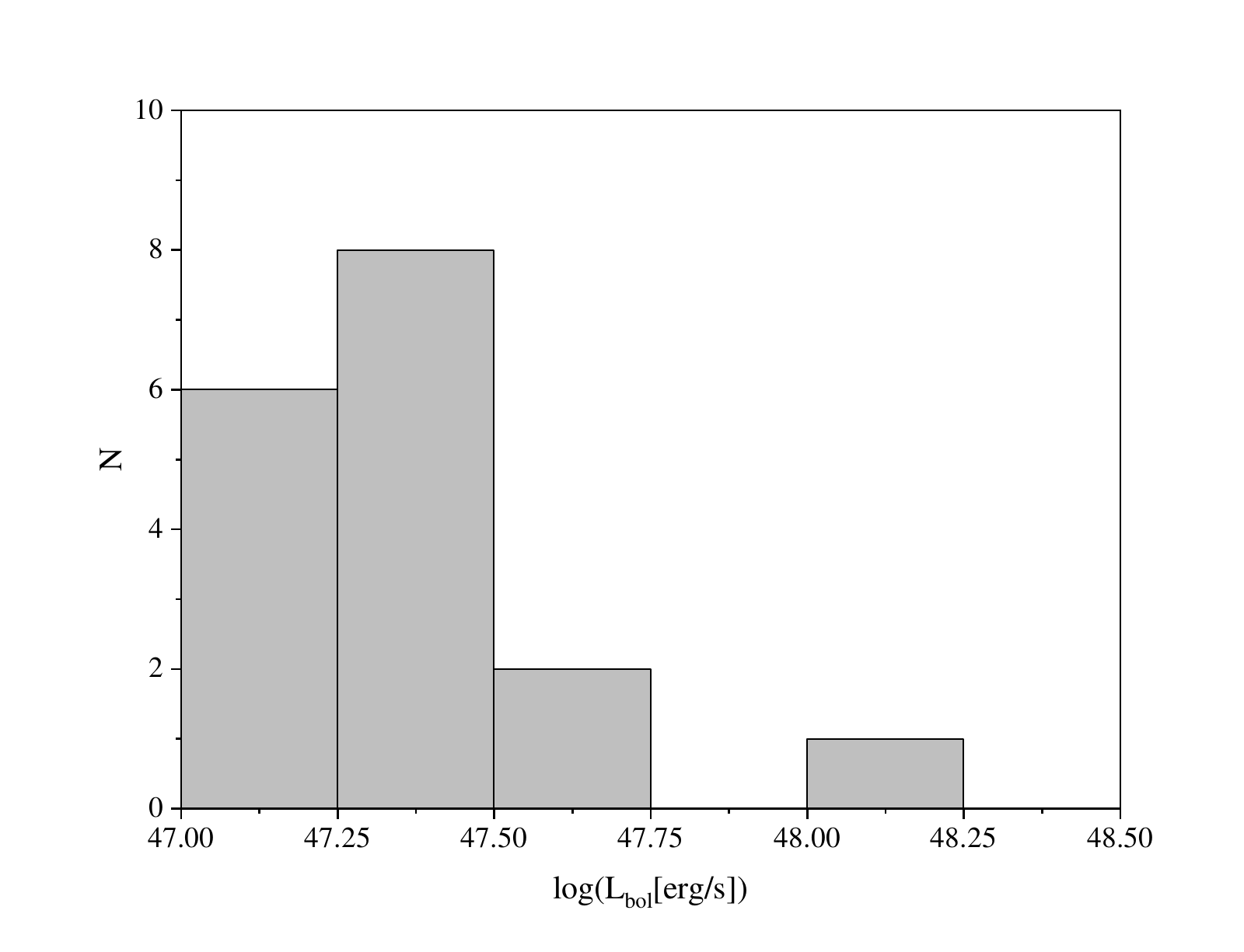}
\caption{Distributions of objects by bolometric luminosity $L_\text{bol}$}
\label{fig_01}
\end{figure}
\begin{figure}[ht!]
\centering
\includegraphics[bb= 60 25 715 530, clip, width=0.7\linewidth]{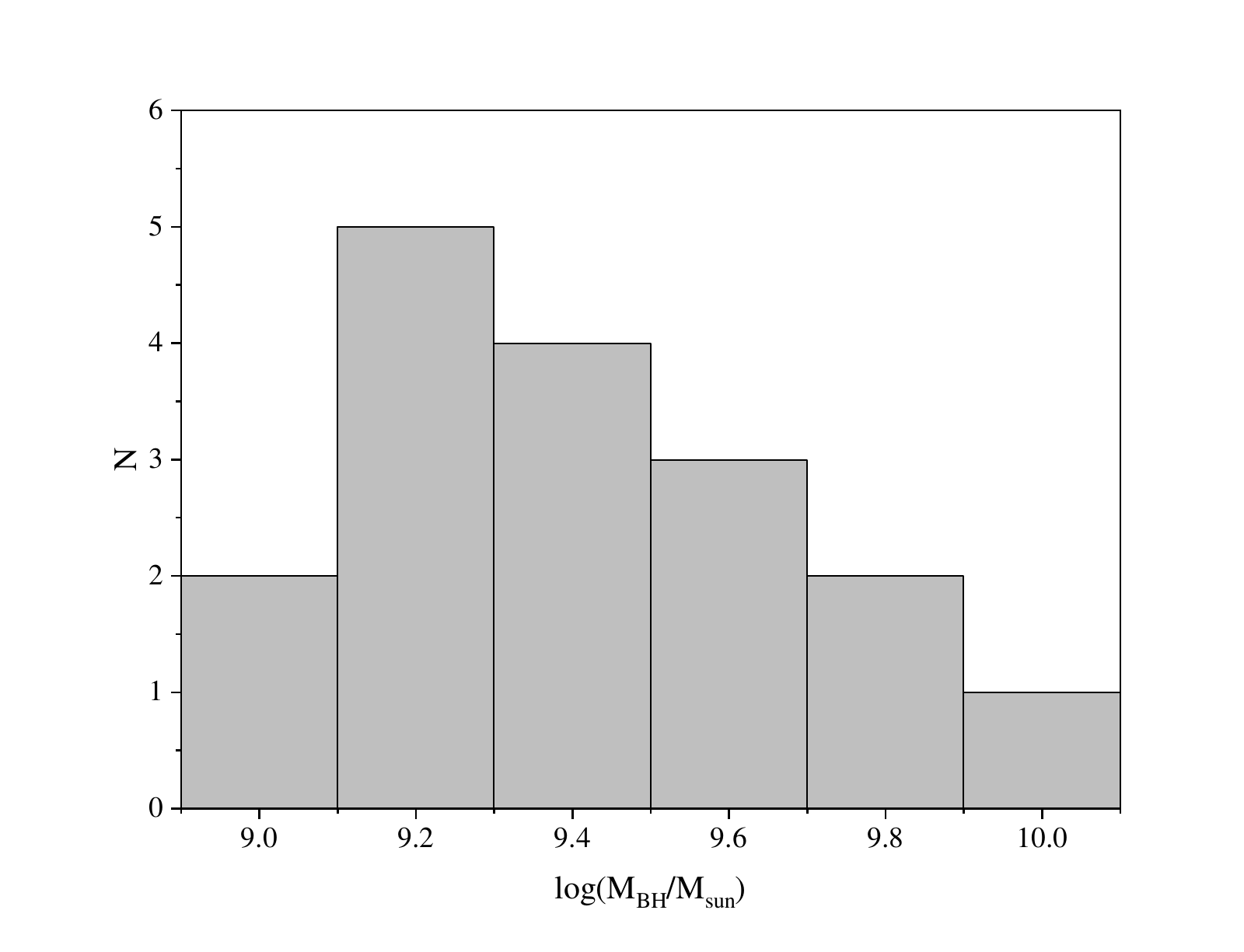}
\caption{Distributions of objects by SMBH mass $M_\text{BH}$}
\label{fig_02}
\end{figure}
\begin{figure}[ht!]
\centering
\includegraphics[bb= 60 25 715 530, clip, width=0.7\linewidth]{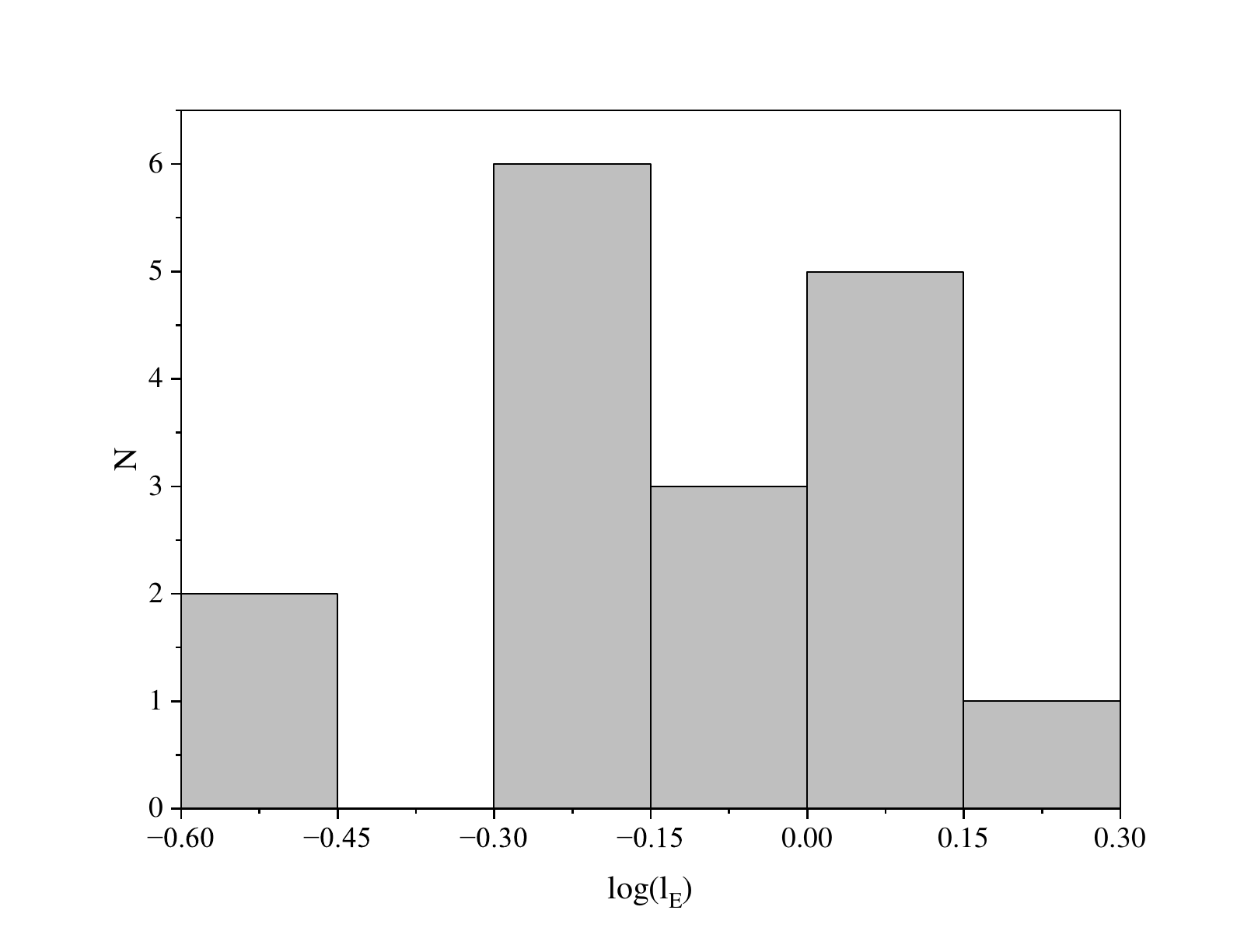}
\caption{Distributions of objects by Eddington ratio $l_\text{E}$}
\label{fig_03}
\end{figure}
\begin{figure}[ht!]
\centering
\includegraphics[bb= 60 25 715 530, clip, width=0.7\linewidth]{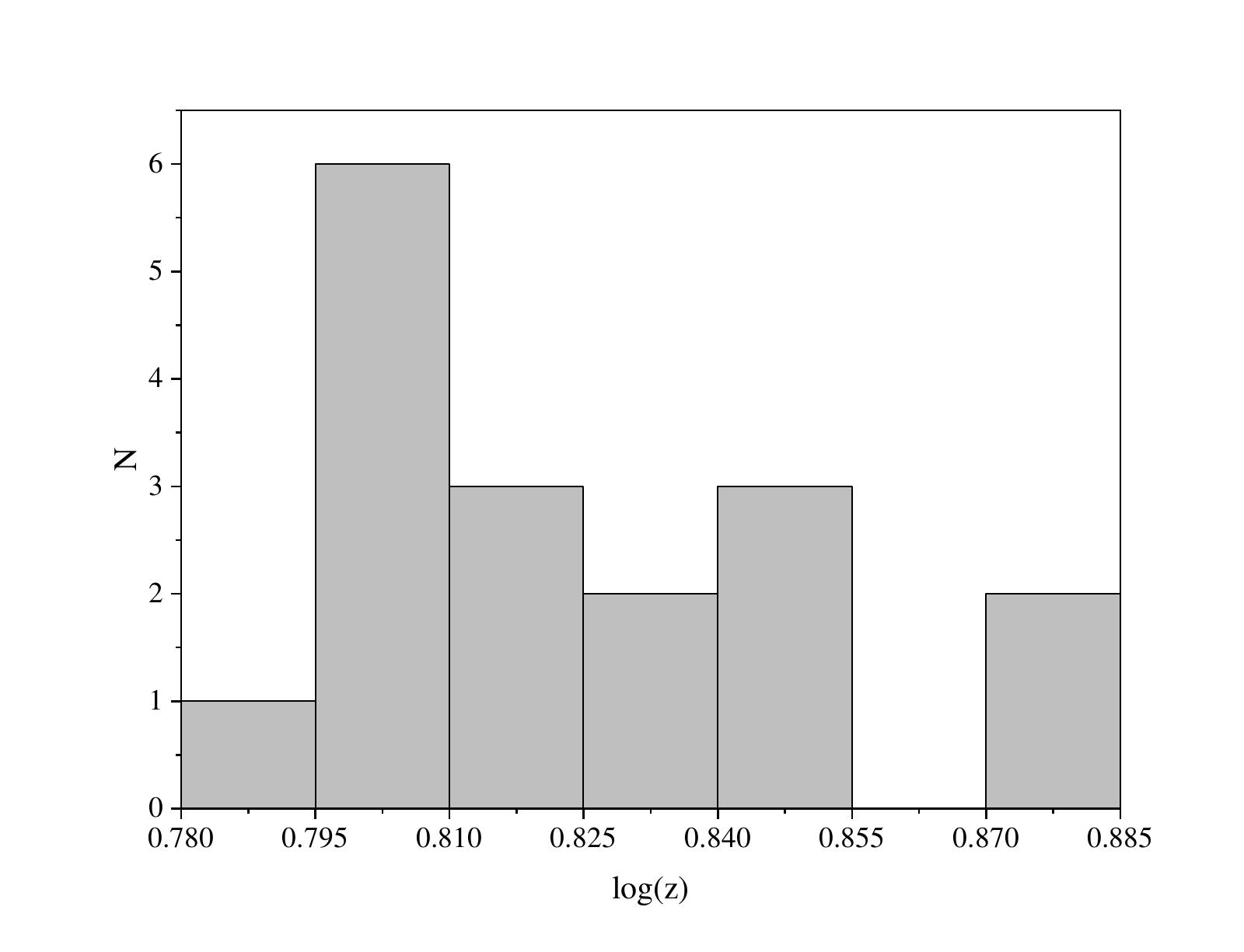}
\caption{Distributions of objects by cosmological redshift $z$}
\label{fig_04}
\end{figure}
We took a sample of 18 distant ($6 < z < 7.5$) ultraluminous QSOs from Ref.~\refcite{zappacosta23} (Table~\ref{table_01}), this sample includes the
"titans" among QSOs, i.e. those powered by the SMBH which had the largest mass assembly over the Universe first Gyr. Object VHS~J0411-0907 from this sample stands out from the others because of its unusually high Eddington ratio 2.57. There are two possible explanations here. Either it has different (from the others) accretion mechanism (and different type of accretion disk), or its SMBH mass or bolometric luminosity values were determined incorrectly. Therefore, we decided to exclude this object from further consideration.

In Ref.~\refcite{zappacosta23}, the uncertainties of the physical parameters we need are not explicitly stated. So we decided to use the following uncertainty values. For cosmological redshift $z$ 0.005 \cite{Fan06}. For luminosities, masses and the Eddington ratio (in logarithmic scale) 0.1 \cite{vandenberk01,shen11}.

We then performed a statistical analysis of the parameters of the remaining 17 objects. Figs.~\ref{fig_01}-\ref{fig_04} show the distributions of objects by bolometric luminosity $L_\text{bol}$, SMBH mass $M_\text{BH}$, Eddington ratio $l_\text{E}$ and cosmological redshift $z$. It can be seen that despite the small number of objects, the SMBH mass distribution has a form somewhat similar to a log-normal distribution (Fig.~\ref{fig_02}). The remaining distributions do not have any clearly defined character (and they do not demonstrate any clearly defined trends), which is due to the small number of objects, as well as the relatively small spread of mass and redshift values.

\section{Spin estimation method}

\begin{figure}
\centering
\includegraphics[bb= 55 35 700 535, clip, width= 0.7\columnwidth]{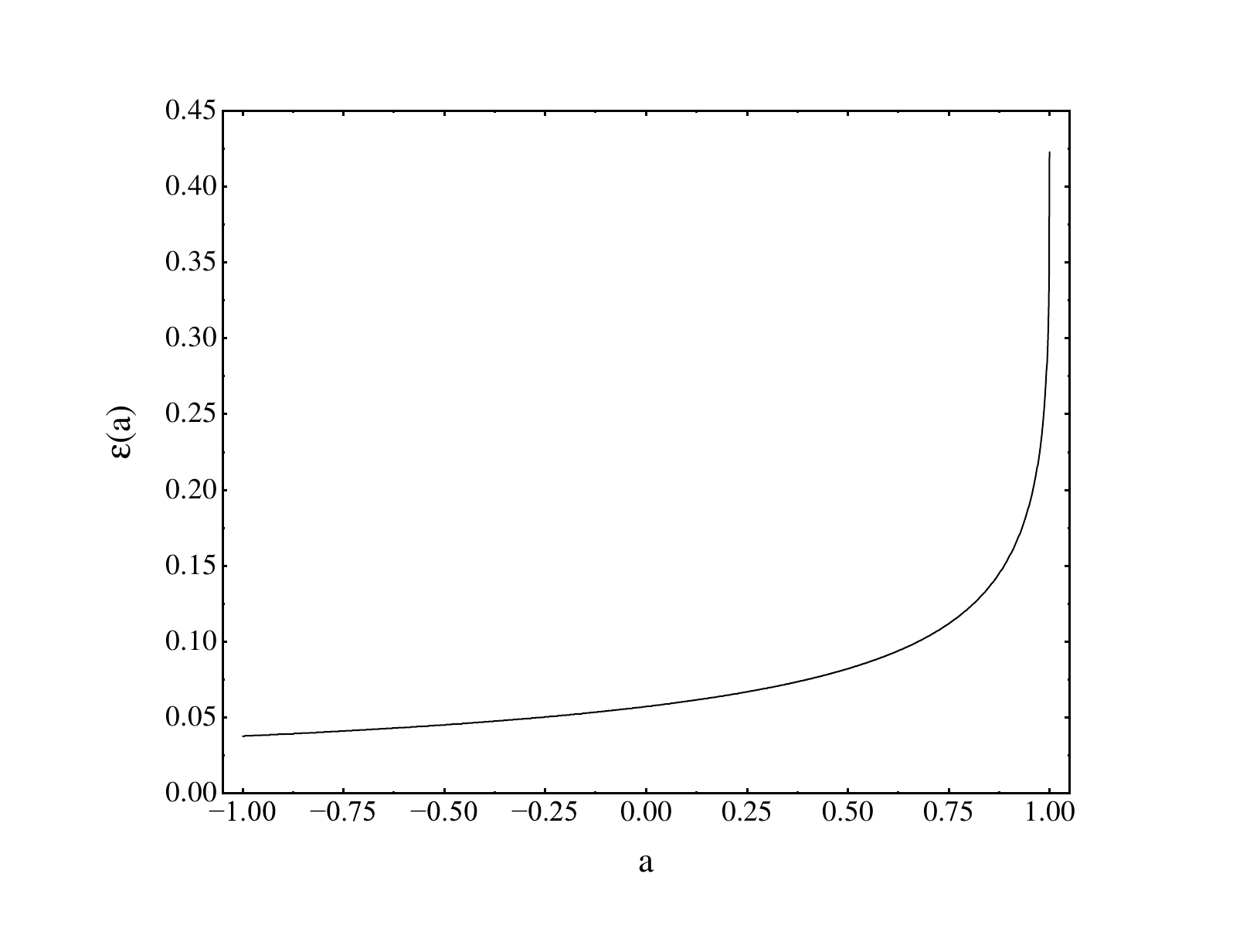}
\caption{Dependence of the radiative efficiency $\varepsilon$ on the spin $a$ {(see Eqs.(\ref{eq01},\ref{eq02}))}.}
  \label{epsilon_fig}
\end{figure}
We can estimate the spin of a black hole by determining the radiative efficiency $\varepsilon(a)$ of its accretion disk, which depends on the black hole's spin \cite{bardeen72,novikov73,krolik07,krolik07b} (see Fig.~\ref{epsilon_fig}). The radiative efficiency is $\varepsilon = L_{\rm bol} / (\dot{M} c^2)$, where $L_{\rm bol}$ is the AGN or quasar bolometric luminosity and $\dot{M}$ is the accretion rate. The radiative efficiency should fall within the range $0.039 < \varepsilon < 0.324$, and the spin $-1.0 < a \leq 0.998$ \cite{thorne74}. A negative value of the spin refers to ''retrograde'' rotation, where accretion disk and SMBH rotate in opposite directions.

The radiative efficiency depends on various parameters: the SMBH mass $M_{\rm BH}$, the inclination angle between the line of sight and the normal to the accretion disk plane $i$, and the bolometric luminosity $L_{\rm bol}$ \cite{davis11,raimundo11,du14,trakhtenbrot14,lawther17}. All of this are based on statistical analyses of observational data and on the use of the Shakura-Sunyaev accretion disk model \cite{shakura73}. Since we are studying extremely remote objects, in this case we consider it most correct to use the model from Ref.~\refcite{trakhtenbrot14}, which is designed specifically for such kind of objects:
\begin{equation}
 \varepsilon \left( a \right) =  0.073 \left(\frac{L_\text{bol}}{10^{46}\text{erg/s}}\right) \left(\frac{L_\text{opt}}{10^{45}\text{erg/s}}  \right)^{-1.5} \times \left(\frac{4400\text{\AA}}{5100\text{\AA}}\right)^{-2} M_8 \mu^{1.5}.
 \label{eq03}
\end{equation}

\noindent Here $L_\text{opt}$ is the optical luminosity (at 4400\AA), $M_8 = \frac{M_\text{BH}}{10^8 M_{\odot}}$ and $\mu = \cos{(i)}$.

Determining bolometric corrections factors that are used for obtaining luminosity value at a certain wavelength from bolometric luminosity is a complex problem. We can find factors in literature that differ 2-3 times \cite{richards06,hopkins07,cheng19,netzer19,duras20}. In this work we decided to use definition of optical luminosity $L_\text{opt}$ from Ref.~\refcite{hopkins07}:
\begin{equation}
  \frac{L_\text{bol}}{L_\text{opt}} = 6.25 \left(\frac{L_\text{bol}}{10^{10} L_\odot}\right)^{-0.37} + 9.0 \left(\frac{L_\text{bol}}{10^{10} L_\odot}\right)^{-0.012},
\end{equation}

\noindent where $L_\odot$ is the luminosity of the Sun.

Reliable determination of inclination angle $i$ from observational data is a complicated and not yet fully solved problem. It is usually simply assumed that $i$ equals to some average constant. We, on the other hand, employed the following approach: for each object, we took the average value of the angle $i = 45^{\circ}$ and if the numerical method did not give a physically meaningful result, then we changed the angles to smaller and larger values with a step size of $5^{\circ}$ until achieving a meaningful outcome.

Determined mass of SMBH depends on inclination angle $i$, since the commonly used method for determining mass uses relation \cite{decarli08}:
\begin{equation}
  M_{\rm BH} = \frac{R_{\rm BLR} V_{\rm BLR}^2}{G},
  \label{eq_M_BH}
\end{equation}

\noindent where $R_{\rm BLR}$ is the accretion disk broad-line region (BLR) scale radius, $V_{\rm BLR}$ is the typical velocity of matter in BLR and $G$ is the gravitational constant. $V_{\rm BLR}$ can be determined through observations, for example, by measuring full width at half maximum (FWHM) of H$\beta$ spectral line:
\begin{equation}
  V_{\rm BLR} = f \times FWHM({\rm H}\beta).
  \label{eq_V_BLR}
\end{equation}

\noindent Here $f$ is coefficient describing the geometry of the accretion disk that can be expressed as \cite{decarli08}:
\begin{equation}
  f = \left(2\sqrt{\left(\frac{H}{R}\right)^2 + \sin^2{i}}\right)^{-1},
  \label{eq_f}
\end{equation}

\noindent where $H/R$ is a ratio of the geometric thickness of the disk to the disk radius. Here we assume that our objects have geometrically thin disks, so $H/R \ll 1$ and
\begin{equation}
  f \approx \frac{1}{2 \sin{i}}.
  \label{eq_f1}
\end{equation}

We can determine $R_{\rm BLR}$ from observations using method from Ref.~\refcite{collin04}:
\begin{equation}
  R_{\rm BLR} = 32.9 \times \left(\frac{L_{5100}}{10^{44}\,{\rm erg/s}}\right)^{0.7},
  \label{eq_R_BLR}
\end{equation}

\noindent where $L_{5100}$ is luminosity at 5100\AA. Here we decided to use for consistency bolometric correction for luminosity at 5100\AA\, from \cite{richards06}: $L_{5100} = L_\text{bol} / 10.3$.

As a rule when determining SMBH mass it is assumed by default that $f = \sqrt{3}/2$ (i.e. $i \approx 35^{\circ}$) \cite{collin04,decarli08}. So, in the process of estimating radiative efficiencies for other values of inclination angle we also for self-consistency estimated new mass values $M^*_\text{BH}$ (Eq.(\ref{eq_M_BH})), assuming that mass from the literature was obtained for $i \approx 35^{\circ}$.

The spin values were determined numerically using the method from Ref.~\refcite{bardeen72} {(see Fig.\ref{epsilon_fig})}:
\begin{equation}
  \varepsilon(a) = 1 - \frac{R_\text{ISCO}^{3/2} - 2 R_\text{ISCO}^{1/2} + |a|}{R_\text{ISCO}^ {3/4}\left(R_\text{ISCO}^{3/2} - 3 R_\text{ISCO}^{1/2} + 2 |a|\right)^{1/2}},
  \label{eq01}
\end{equation}

\noindent where $R_\text{ISCO}$ is the radius of the innermost stable circular orbit that strongly depends on the spin:
\begin{equation}
  \begin{array}{l}
   R_\text{ISCO}(a) = \\
   = 3 + Z_2 \pm [(3 - Z_1)(3 + Z_1 + 2 Z_2)]^{1/2},\\
   Z_1 = 1 + (1 - a^2)^{1/3}\left[(1 + a)^{1/3} + (1 - a)^{1/3}\right],\\
   Z_2 = (3 a^2 + Z_1^2)^{1/2}.
  \end{array}
  \label{eq02}
\end{equation}

\noindent Here ''-'' is corresponds to $a \geq 0$, and ''+'' to $a < 0$.

\section{Analysis of the estimated values}

\begin{table}[ht!]
\tbl{Results of our estimations of the values of radiative efficiency $\varepsilon$, spin $a$, inclination angle $i$ and SMBH mass $M^*_\text{BH}$}
{\begin{tabular}{lcccc}
\toprule
Object & $i$[deg] & $\log{\frac{M^*_\text{BH}}{M_\odot}}$ & $\varepsilon$ & $a$\\
\colrule
 ULAS  J1342+0928 & 45 &  8.96$\pm$0.10 &  0.105$_{-0.038}^{+0.059}$ &  0.716$_{-0.442}^{+0.200}$\\
  QSO  J1007+2115 & 45 &  9.24$\pm$0.10 &  0.174$_{-0.063}^{+0.098}$ &  0.934$_{-0.182}^{+0.058}$\\
 ULAS  J1120+0641 & 45 &  9.47$\pm$0.10 &  0.296$_{-0.106}^{+0.025}$ &  0.996$_{-0.046}^{+0.002}$\\
 DELS  J0038-1527 & 45 &  9.20$\pm$0.10 &  0.147$_{-0.053}^{+0.083}$ &  0.884$_{-0.248}^{+0.096}$\\
  DES  J0252-0503 & 45 &  9.21$\pm$0.10 &  0.203$_{-0.073}^{+0.114}$ &  0.964$_{-0.128}^{+0.034}$\\
 VDES  J0020-3653 & 45 &  9.30$\pm$0.10 &  0.238$_{-0.085}^{+0.083}$ &  0.984$_{-0.088}^{+0.014}$\\
 VDES  J0244-5008 & 45 &  9.14$\pm$0.10 &  0.158$_{-0.057}^{+0.089}$ &  0.908$_{-0.218}^{+0.078}$\\
  PSO J231.6-20.8 & 50 &  9.49$\pm$0.10 &  0.265$_{-0.095}^{+0.056}$ &  0.992$_{-0.064}^{+0.006}$\\
  PSO J036.5+03.0 & 50 &  9.48$\pm$0.10 &  0.253$_{-0.091}^{+0.068}$ &  0.988$_{-0.074}^{+0.010}$\\
 VDES  J0224-4711 & 45 &  9.42$\pm$0.10 &  0.199$_{-0.071}^{+0.111}$ &  0.960$_{-0.136}^{+0.038}$\\
  PSO     J011+09 & 45 &  9.21$\pm$0.10 &  0.203$_{-0.073}^{+0.114}$ &  0.964$_{-0.128}^{+0.034}$\\
 SDSS  J1148+5251 & 55 &  9.67$\pm$0.10 &  0.247$_{-0.089}^{+0.074}$ &  0.986$_{-0.078}^{+0.012}$\\
  PSO J083.8+11.8 & 45 &  9.38$\pm$0.10 &  0.286$_{-0.103}^{+0.035}$ &  0.996$_{-0.052}^{+0.002}$\\
 SDSS  J0100+2802 & 50 & 10.03$\pm$0.10 &  0.296$_{-0.106}^{+0.025}$ &  0.996$_{-0.044}^{+0.002}$\\
ATLAS     J025-33 & 50 &  9.56$\pm$0.10 &  0.282$_{-0.101}^{+0.039}$ &  0.994$_{-0.052}^{+0.004}$\\
CFHQS  J0050+3445 & 55 &  9.61$\pm$0.10 &  0.303$_{-0.109}^{+0.018}$ &  0.998$_{-0.042}^{+0.000}$\\
ATLAS     J029-36 & 60 &  9.70$\pm$0.10 &  0.270$_{-0.097}^{+0.051}$ &  0.992$_{-0.060}^{+0.006}$\\
\botrule
\end{tabular} \label{table_02}}
\end{table}

\begin{figure}[ht!]
\centering
\includegraphics[bb= 60 25 715 530, clip, width=0.7\linewidth]{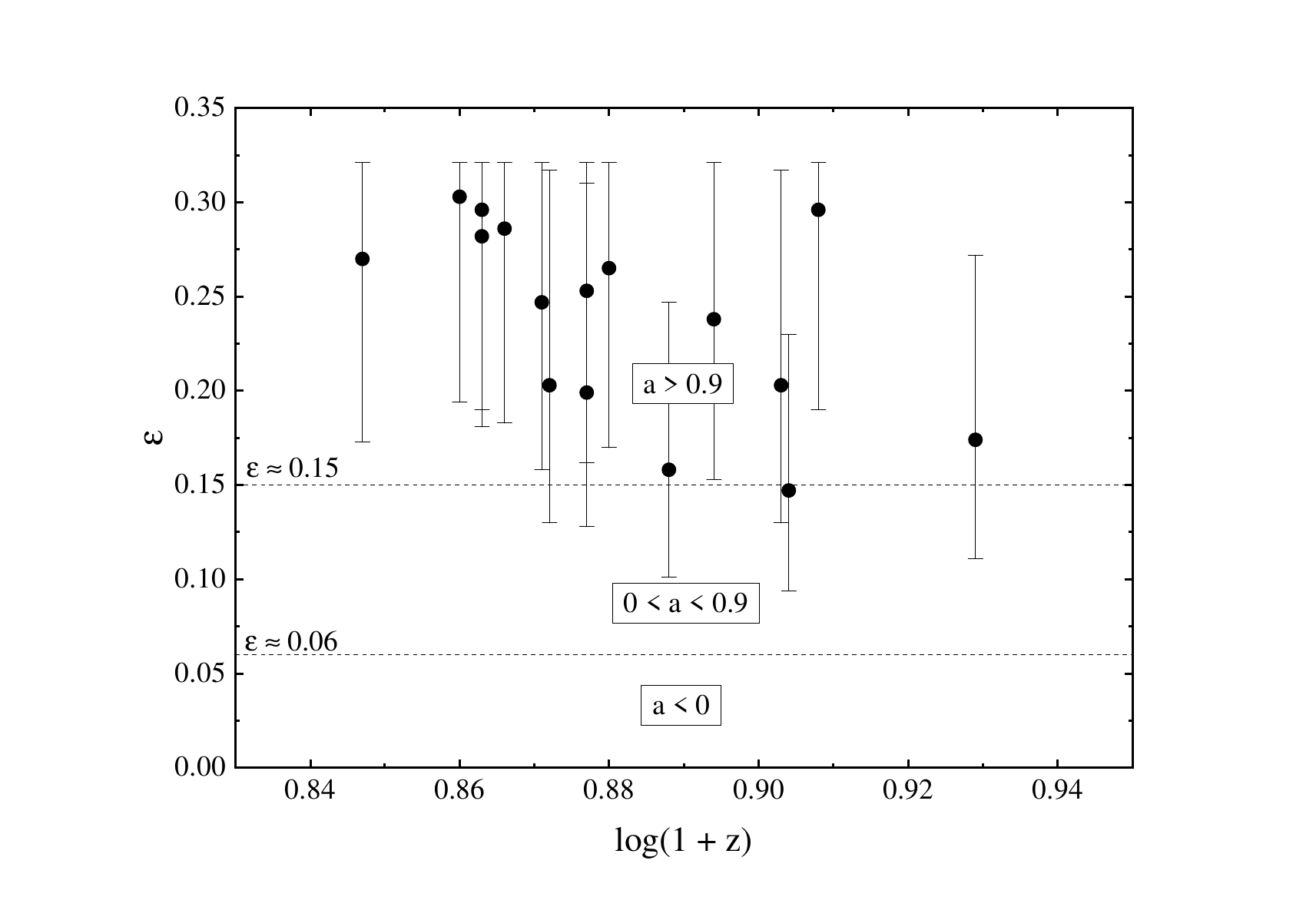}
\caption{Dependence of the estimated radiative efficiency $\varepsilon$ on cosmological redshift $z$}
\label{fig_06}
\end{figure}
\begin{figure}[ht!]
\centering
\includegraphics[bb= 60 25 715 530, clip, width=0.7\linewidth]{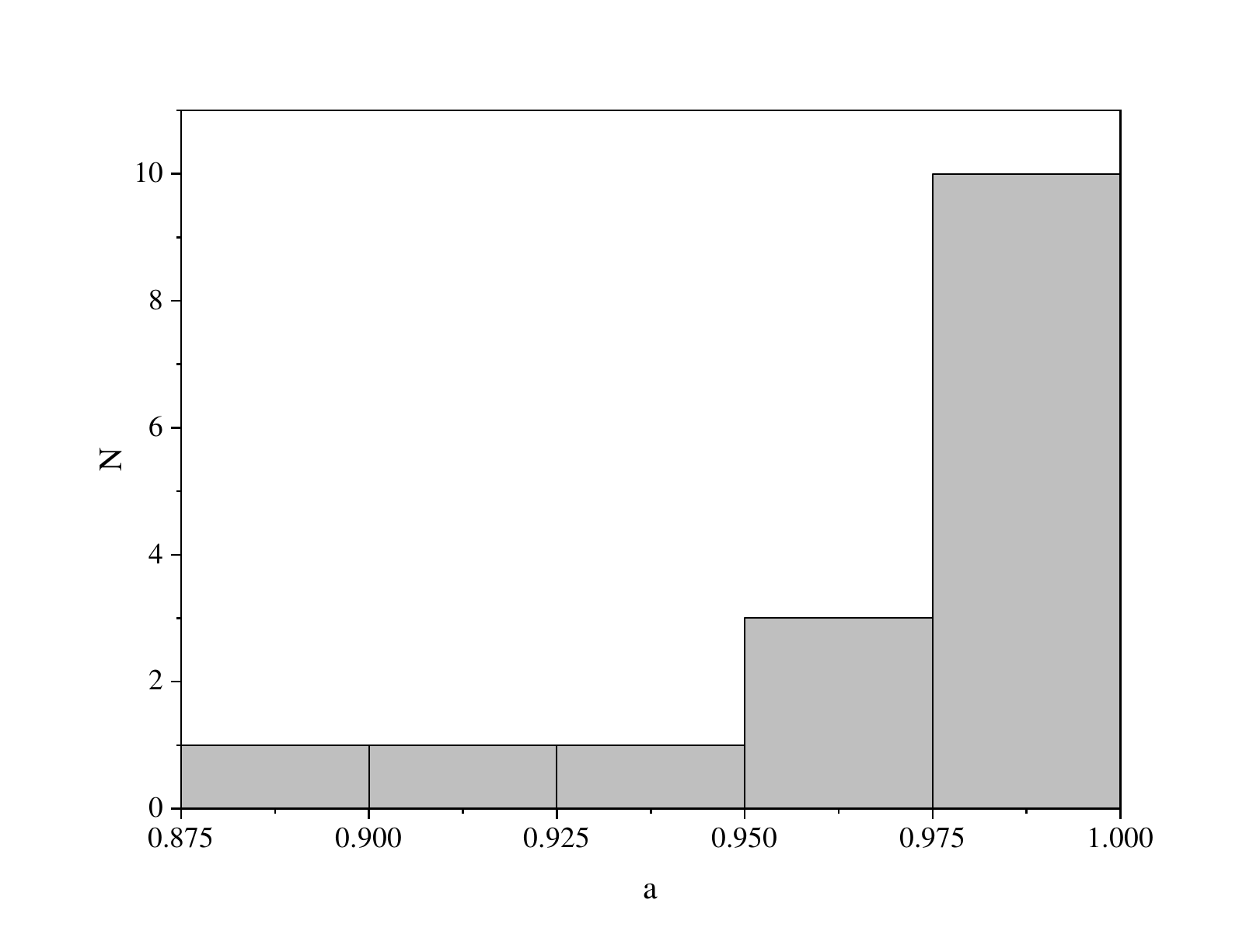}
\caption{Distribution of the estimated spin $a$}
\label{fig_07}
\end{figure}
\begin{figure}[ht!]
\centering
\includegraphics[bb= 60 25 715 530, clip, width=0.7\linewidth]{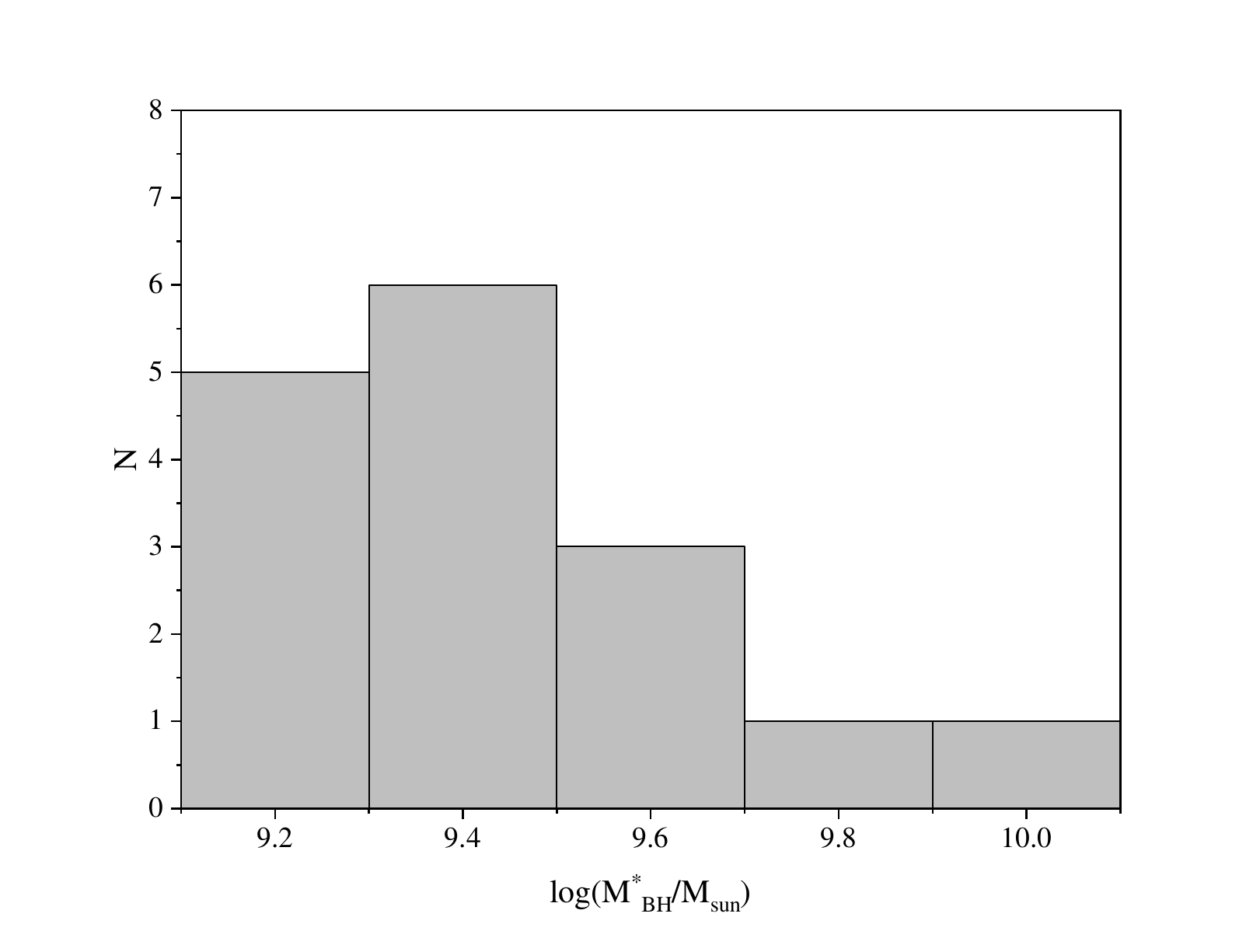}
\caption{Distribution of the estimated SMBH mass $M^*_\text{BH}$}
\label{fig_08}
\end{figure}
\begin{figure}[ht!]
\centering
\includegraphics[bb= 60 25 715 530, clip, width=0.7\linewidth]{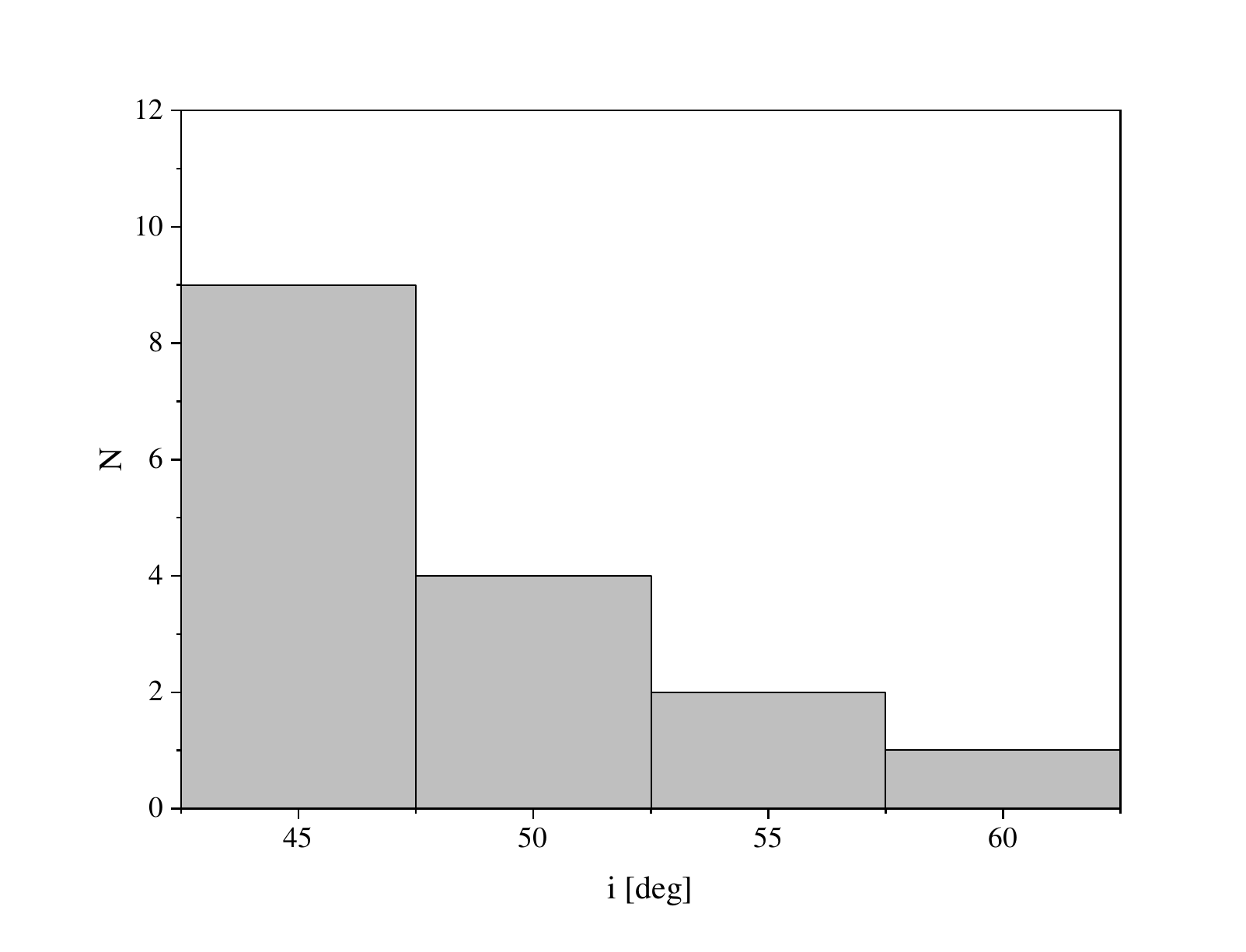}
\caption{Distribution of the estimated inclination angle $i$}
\label{fig_09}
\end{figure}
\begin{figure}[ht!]
\centering
\includegraphics[bb= 60 25 715 530, clip, width=0.7\linewidth]{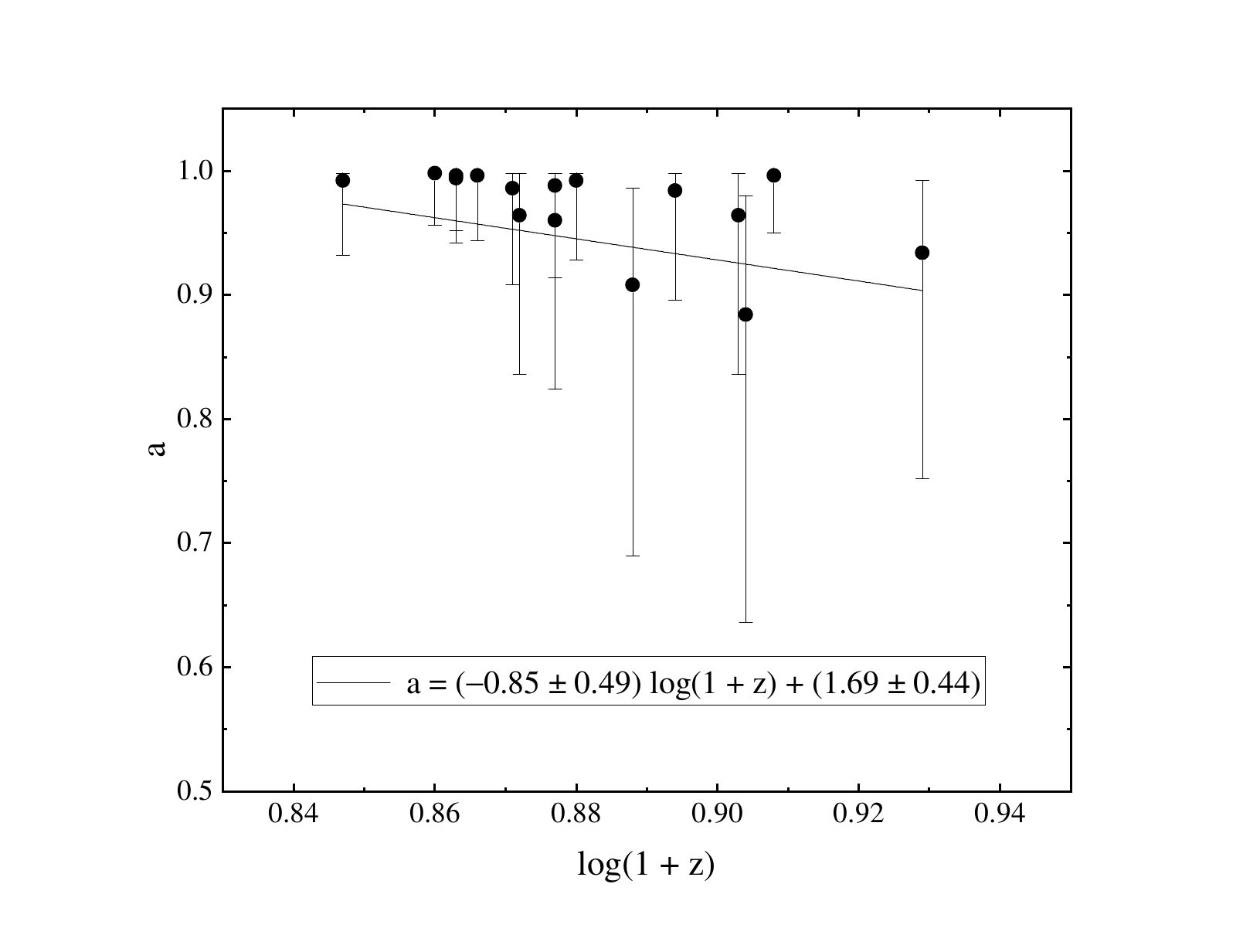}
\caption{Dependence of the estimated spin $a$ on cosmological redshift $z$}
\label{fig_10}
\end{figure}
\begin{figure}[ht!]
\centering
\includegraphics[bb= 60 25 715 530, clip, width=0.7\linewidth]{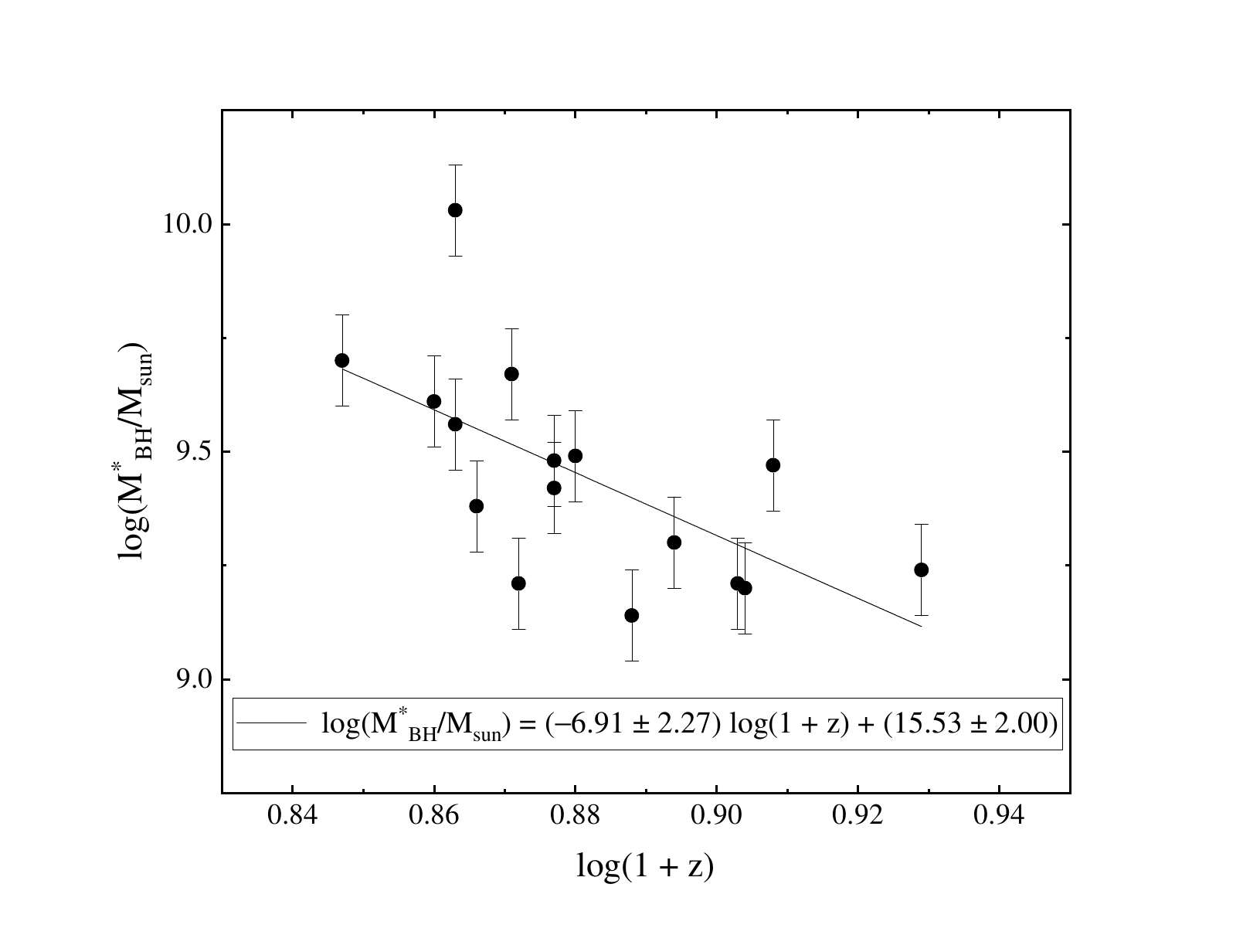}
\caption{Dependence of the estimated SMBH mass $M^*_\text{BH}$ on cosmological redshift $z$}
\label{fig_11}
\end{figure}
\begin{figure}[ht!]
\centering
\includegraphics[bb= 60 25 715 530, clip, width=0.7\linewidth]{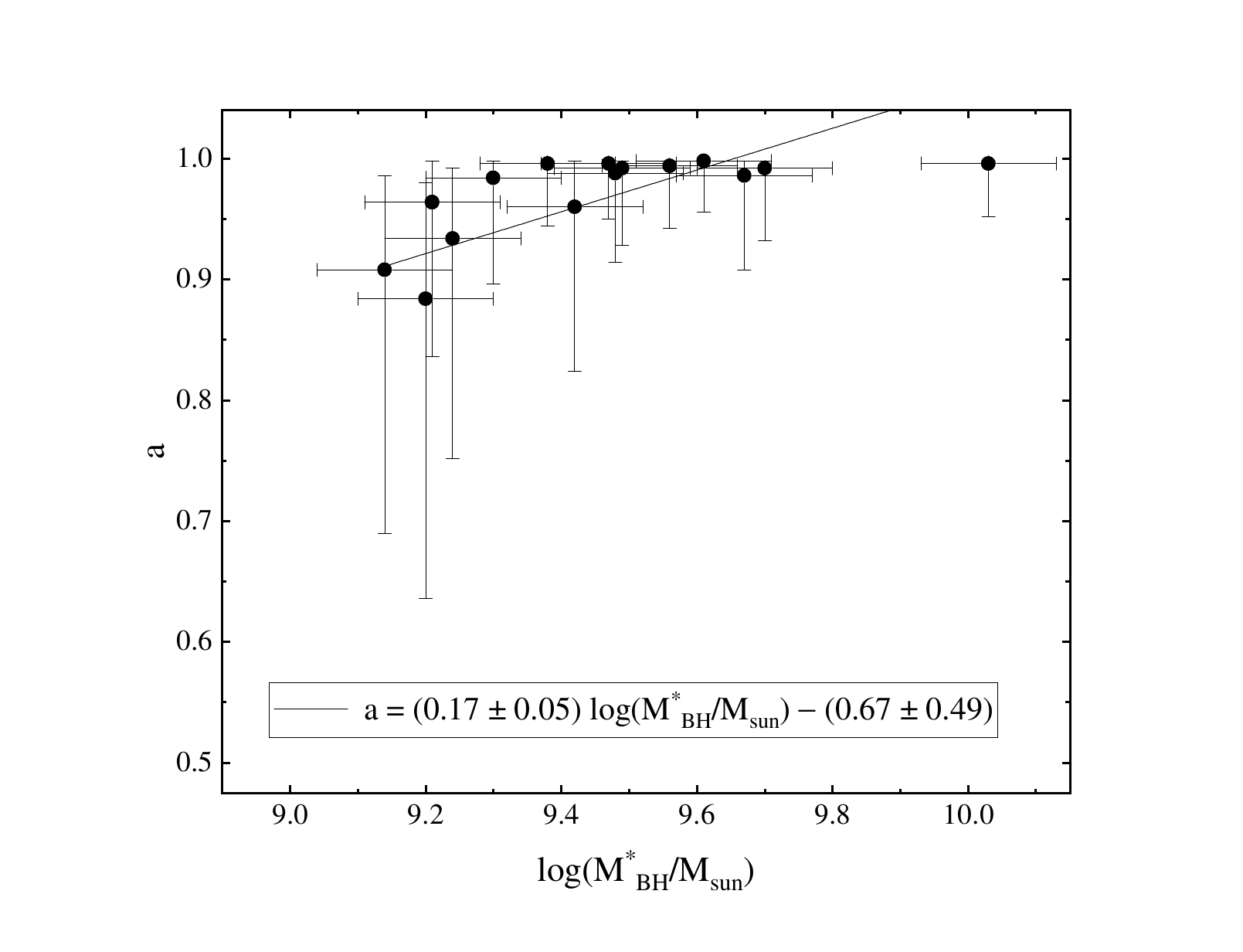}
\caption{Dependence of the estimated spin $a$ on SMBH mass $M^*_\text{BH}$}
\label{fig_12}
\end{figure}

Table~\ref{table_02} shows the results of our estimations of the value of radiative efficiency, spin, inclination angle and SMBH mass for our objects. It can be seen that spin values are mostly greater that 0.9, which is expected because statistically most known SMBH in AGNs and quasars have similar spin value \cite{trakhtenbrot14,daly19,reynolds21,azadi23}. Due to the calculation method inclination angle values in this case should be considered as a lower estimate. It should be noted that we do not indicate the errors in the inclination angle values, since formally we did not receive it as a direct result of the calculations or observations, but we set the exact value ourselves. It is also clear from the Table~\ref{table_02} that for object ULAS~J1342+0928 the spin value is obtained with a significant error, much greater than for all other objects. From which we can conclude that the calculation method we use is most likely not suitable for this object. So we decided to remove this object from further consideration.

Figs.~\ref{fig_06}-\ref{fig_12} show the statistical analysis and correlation between the estimates parameters.

Fig.~\ref{fig_06} demonstrates the dependence of estimated radiative efficiency on cosmological redshift. The nature of the dependence of the spin on the radiative efficiency (Fig.~\ref{epsilon_fig}) indicates that there are three reasonably possible ranges of values for the spin: for $\varepsilon \lesssim 0.06$, $a < 0$, for $0.06 \lesssim \varepsilon \lesssim 0.15$, $0 < a < 0.9$, and for $\varepsilon \gtrsim 0.15$, $a > 0.9$. We can see from Fig.~\ref{fig_06} that the vast majority of points (taking into account the error margins) lie in the area corresponding to $a > 0.9$.

Fig.~\ref{fig_07} shows the spin distribution. It has a characteristic appearance, similar to ones obtained for other types of AGNs and quasars: distant quasars \cite{trakhtenbrot14}, Seyfert galaxies \cite{afanasiev18,piotrovich22} and ''red'' quasars \cite{piotrovich24}. It should be noted that our objects have a slightly higher average spin value, which is most likely due to the fact that our sample contains only ultraluminous objects (selection bias), whose high luminosity is associated, among other things, with a high radiative efficiency, which in turn depends on spin.  From this we can assume that the mechanism for increasing the angular momentum of SMBHs of these extremely distant objects is similar to the mechanism for closer ones.

Fig.~\ref{fig_08} shows the estimated SMBHs mass distribution. It has a shape similar to the log-normal one of initial mass distribution form \cite{zappacosta23} (Fig.~\ref{fig_02}) however the peak of the distribution now falls at a slightly higher mass value. This is because we took larger average inclination angles in our model than are usually assumed by default when calculating mass. As a result, our estimated masses are slightly larger on average.

Fig.~\ref{fig_09} demonstrates the estimated inclination angles distribution. Generally speaking, the actual values of these angles should be distributed randomly. Thus, in principle, given the small number of objects and the fact that these values are lower bounds, such a distribution looks plausible.

Fig.~\ref{fig_10} displays the dependence of estimated spin on cosmological redshift. (It should be noted that redshift $6 < z < 7.5$ corresponds to time from the Big Bang as $9.4\times 10^7 \text{years} > t > 7\times 10^7\text{years}$.) There is moderate anti-correlation between parameters (Pearson correlation coefficient is -0.43). Linear fitting gives us: $a = (-0.85 \pm 0.49) \log(1 + z) + (1.69 \pm 0.44)$, i.e. spin increases noticeably over time. From this we can conclude that in this early quasars the growth of SMBHs mass should occur mainly due to disk accretion with high accretion rate, which very effectively increases the spin.

Fig.~\ref{fig_11} shows the dependence of estimated SMBHs mass on cosmological redshift. There is strong anti-correlation between parameters (Pearson correlation coefficient is -0.63). Linear fitting gives us: $\log{(M^*_\text{BH} / M_\odot)} = (-6.91 \pm 2.27) \log(1 + z) + (15.53 \pm 2.00)$, i.e. mass rapidly grows with time, which confirms our conclusion about the high accretion rate.

Fig.~\ref{fig_12} demonstrates the dependence of estimated spin on SMBHs mass. We can see strong correlation between parameters (Pearson correlation coefficient is 0.67). Linear fitting gives us: $a = (0.17 \pm 0.05) \log{(M^*_\text{BH} / M_\odot)} - (0.67 \pm 0.49)$, i.e. we clearly see the growth of SMBH spin with mass. Let us note that in our previous works \cite{piotrovich23,piotrovich24} the dependence of spin on mass (for closer objects) was stronger. This is most likely due to the selection bias: our initial sample consist of ultraluminous objects with high average spin values.

\section{Conclusion}

We estimated spin, inclination angle and corresponding SMBH mass values for sample of extremely distant ($6 < z < 7.5$) ultraluminous quasars. The estimated spin values are on average greater that 0.9 and the spin distribution has a characteristic appearance, similar to ones obtained for other types of AGNs and quasars: distant quasars \cite{trakhtenbrot14}, Seyfert galaxies \cite{afanasiev18,piotrovich22} and ''red'' quasars \cite{piotrovich24}. The estimated SMBHs mass distribution has a shape similar to the log-normal one of initial mass distribution form Ref.~\refcite{zappacosta23} however the peak of the distribution now falls at a slightly higher mass value.

The dependence of estimated spin on cosmological redshift shows moderate anti-correlation between parameters in the form of $a = (-0.85 \pm 0.49) \log(1 + z) + (1.69 \pm 0.44)$, i.e. spin noticeably grows with time, from which we can assume that in this early quasars the growth of SMBHs mass should occur mainly due to disk accretion with high accretion rate, which very effectively increases the spin.

The dependence of estimated SMBHs mass on cosmological redshift shows strong anti-correlation between parameters in the form of $\log{(M^*_\text{BH} / M_\odot)} = (-6.91 \pm 2.27) \log(1 + z) + (15.53 \pm 2.00)$, which confirms our conclusion about the high accretion rate.

The dependence of estimated spin on SMBHs mass demonstrates strong correlation of the parameters in the form of $a = (0.17 \pm 0.05) \log{(M^*_\text{BH} / M_\odot)} - (0.67 \pm 0.49)$, which confirms our assumption that the growth of SMBHs mass should occur mainly due to disk accretion.

\section*{Acknowledgments}

We are grateful to the Reviewer for very useful comments.

\bibliographystyle{ws-ijmpd}
\bibliography{mybibfile}

\end{document}